\documentclass{journalA4}

\pdfoutput=1

\newcommand{\etal}{\emph{et~al.}}
\newcommand{\f}{Fr\'echet }
\newcommand{\df}{\ensuremath d_\textnormal{F}}
\newcommand{\R}{\ensuremath \mathbb{R}}

\renewcommand{\paragraph}[1]{\smallskip\noindent{\bf\sffamily #1.}}

\title{Map Matching with Simplicity Constraints}

\author{Wouter Meulemans\thanks{Dept. of Mathematics and Computer Science, TU Eindhoven, The Netherlands, {\tt w.meulemans@tue.nl}.}}

\begin{document}
\maketitle

\begin{abstract}
We study a map matching problem, the task of finding in an embedded graph a path that has low distance to a given curve in $\R^2$.
The \f distance is a common measure for this problem.
Efficient methods exist to compute the best path according to this measure.
However, these methods cannot guarantee that the result is simple (i.e. it does not intersect itself) even if the given curve is simple.
In this paper, we prove that it is in fact NP-complete to determine the existence a simple cycle in a planar straight-line embedding of a graph that has at most a given \f distance to a given simple closed curve.
We also consider the implications of our proof on some variants of the problem.
\end{abstract}

\section{Introduction}\label{sec:introduction}

Map matching is the task of finding in an embedded graph a path that resembles a give curve.
One of the main applications is finding driven routes in a road network based on a sequence of GPS positions \cite{newson2009,wenk2006}.
The \emph{\f distance} is a prominent similarity measure for this problem.
This distance can be computed efficiently for two polygonal curves \cite{alt1995,buchin2012} and for a polygonal curve and an embedded graph \cite{alt2003,wenk2006}.
However, these methods cannot guarantee that the result is a simple path.
Here we focus on the map matching problem under the \f distance with such simplicity constraints.
To properly define the problem, we first introduce some terminology and notation.

\paragraph{Preliminaries}
A curve $C$ is a continuous function to $\R^2$.
An \emph{open curve} uses $[0,1]$ as domain, $C \colon [0,1] \rightarrow \R^2$.
A \emph{closed curve} uses the unit circle $S^1$ as domain, $C \colon S^1 \rightarrow \R^2$.
A curve is called \emph{simple} if it does not overlap itself, i.e. $C(t) \neq C(t')$ for any distinct $t$ and $t'$ from the domain.

Let $C_1$ and $C_2$ denote two open or two closed curves, and $D$ their domain (i.e. $[0,1]$ for open curves; $S^1$ for closed curves).
Let $\Psi$ be the set of orientation-preserving homeomorphisms for $D$.
Then the \emph{\f distance} between $C_1$ and $C_2$ is defined as
\[ \df(C_1,C_2) = \inf_{\psi \in \Psi} \max_{t \in D}\{ |C_1(t) - C_2(\psi(t))| \}, \]
where $|x - y|$ denotes the Euclidean distance between points $x$ and $y$ in $\R^2$.
In other words, it is the maximal distance obtained by a monotonous ``reparameterization'' of $C_2$.
The \emph{weak \f distance} is a variant characterized by allowing non-monotonous reparameterizations.

We refer to a planar straight-line embedding of a graph in $\R^2$ as a \emph{network}.
A path or cycle in the network is called \emph{simple} if it visits each vertex at most once.
A path or cycle is also a curve and thus the \f distance is defined for these.

\paragraph{Problem statement and contributions}
Let $G$ be a network and let $C$ be a simple closed curve.
We study the problem of determining the existence of a simple cycle $P$ in $G$ with $\df(C,P) \leq \varepsilon$.
This problem is invariant under scaling; hence we assume without loss of generality that $\varepsilon$ is $1$.

We prove that this problem is NP-complete (Section~\ref{sec:nphard}).
The problem also admits a number of variants.
In Section~\ref{sec:corollaries}, we consider the implications of our proof on some of these variants.

\paragraph{Related work}
A large number of papers are concerned with map matching problems, e.g. \cite{alt2003,brakatsoulas2005,newson2009,wenk2006}.
Here we focus on establishing the complexity of variants under the \f distance.
Alt~\etal~\cite{alt2003} describe an algorithm that decides whether a path exists in $O(mn \log n)$ time, where $m$ and $n$ are the number of vertices of the (polygonal) curve and the network respectively.
Though ``U-turns'' can be avoided, no general simplicity guarantees are possible.
Similarly, the decision problem for the weak \f distance can be solved in $O(mn)$ time \cite{brakatsoulas2005}.
The algorithmic complexity for map matching with simplicity constraints (under any reasonable metric) seems to be open.
Studying a slightly different problem, Sherette and Wenk \cite{sherette2013} show that it is NP-hard to determine the existence of a simple curve on a 2D surface with holes or in 3D.
However, their proof does not extend to the case where the input curve is simple.

\section{Simple map matching is NP-complete}\label{sec:nphard}

In this section, we prove the following theorem.
\begin{theorem}\label{thm:main}
Let $G$ be a planar straight-line embedding of a graph (a network) and let $C$ be a simple closed curve.
It is NP-complete to decide whether $G$ contains a simple cycle $P$ with $\df(C,P) \leq 1$.
\end{theorem}
The problem is in NP since the \f distance can be computed in polynomial time \cite{alt1995} and it is straightforward to check simplicity.
In the remainder of the section, we prove that the problem is also NP-hard.

\begin{figure}[t]
\centering
\includegraphics{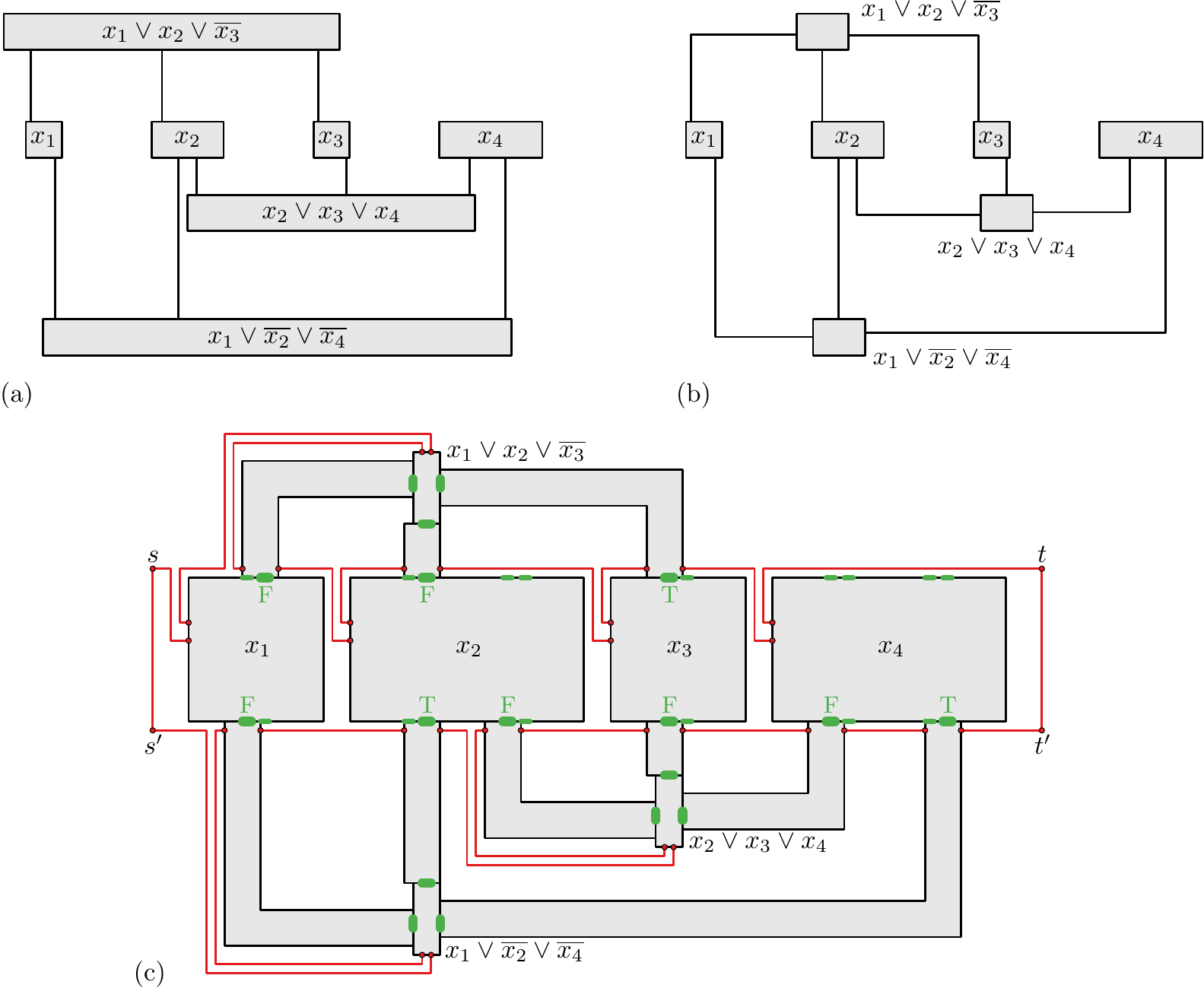}
\caption{(a) 3-SAT formula given as a rectilinear drawing \cite{knuth1992}. (b) The same embedding but with bends and a fixed width for clauses. (c) Result of the construction. Each gray block represents a gadget. The red lines connect the various gadgets to obtain a single closed curve.}
\label{fig:input}
\end{figure}

\paragraph{Planar 3-SAT}
We give a reduction from planar 3-SAT.
For any planar 3-SAT formula $F$, we show how to construct network $G$ and simple closed curve $C$ such that $G$ contains a simple cycle $P$ with $\df(C,P) \leq 1$ if and only if $F$ is satisfiable.
Lichtenstein~\cite{lichtenstein1982} showed that planar 3-SAT is NP-hard.
Knuth and Raghunathan~\cite{knuth1992} proved that it remains NP-hard, even if a specific rectilinear embedding is given.
In this embedding each variable and clause occupies a rectangular area; the variables lie on a horizontal line; and all links relating variables to clauses are strictly vertical (see Figure~\ref{fig:input}(a)).
However, this is not exactly the embedding we use for construct the network and curve.
A requirement of our reduction is that the clauses have a small fixed width and thus cannot be stretched horizontally.
We observe that a single bend for the two ``outer'' edges of a clause are sufficient to ensure this (see Figure~\ref{fig:input}(b)).

As is common in reductions from planar 3-SAT, we define a number of \emph{gadgets}.
We define \emph{variable gadgets}, \emph{clause gadgets}, and \emph{propagation gadgets} which represents the variables, clauses, and edges of $F$ respectively.
Based on these, we construct a network and a simple closed curve which together represent $F$.
The result of this construction is illustrated in Figure~\ref{fig:input}(c).
The reduction is split into two parts.
In Section~\ref{ssec:overview}, we present the reduction based on a specification of gadgets.
In Section~\ref{ssec:gadgets}, we provide the details for each gadget.

\subsection{Proof with gadget specifications}\label{ssec:overview}

Each gadget specifies part of the network and part of the closed curve.
We call these the \emph{local} network and \emph{local} curve of a gadget.
The gadgets interact via vertices and edges shared by their local networks.
There is no interaction based on the curve itself.

For now, we abstract from the details of the local network and curve.
We first give only specifications for the gadgets; we describe their desired behavior.
Based on this specification, we complete the reduction.
In Section~\ref{ssec:gadgets}, we give the construction for each gadget.

For a cycle to exist in the total network, some simple path in the local network must have \f distance at most 1 to the local curve.
Choosing a certain path ``claims'' its vertices and edges: the shared vertices and edges can no longer be used by another gadget.
This results in \emph{pressure} on the other gadget to choose a different path.
A gadget has a number of pressure \emph{ports}.
These ports correspond to an edge (and its vertices) of the local network that may be shared with another gadget.
A port may \emph{receive} pressure, indicating that the shared edge or its vertices may not be used in the gadget.
Similarly, it may \emph{give} pressure, indicating that the shared edge or its vertices may not be used by other gadgets.

The local curves must be joined carefully, ensuring that the total curve is a simple closed curve.
Each gadget has two curve \emph{gates} that correspond to the endpoints of the local curve.
Later, we show how to connect these gates such that the total curve is indeed a simple closed curve.
The total network is the union of all local networks and some additional edges used to connect the gates.

In the following paragraphs, we give the specifications for the three gadgets.
Each specification consists of the following items:
\begin{itemize}
\item its behavior in terms of its ports;
\item a rectilinear bounding polygon that contains the local network and local curve;
\item the placement of its two gates and its ports.
\end{itemize}
We give specifications for clauses and edges occurring above the variables in the embedded formula.
Gadgets for clauses and edges below are defined analogously.

For each of the gadgets, we also give a visual representation on an integer grid.
The bounding polygon is given with a black outline; the ports are represented with thick green lines; the gates are represented with red dots.

\begin{figure}[h]
\centering
\includegraphics{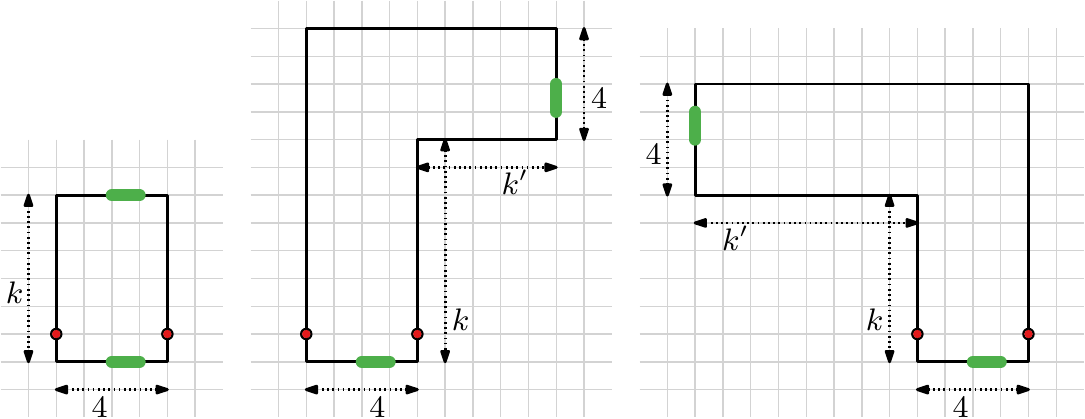}
\caption{Specification of a propagation gadget. The bounding polygon is given in black; the red dots represent gates; the green segments indicate ports. (left) No bend, $k \geq 6$. (middle) A right bend, $k, k' \geq 5$. (right) A left bend, $k, k' \geq 5$.}
\label{fig:spec_propagation}
\end{figure}

\paragraph{Propagation gadget}
Each edge of the embedded formula is represented by a propagation gadget.
The propagation gadget can be considered as a ``thick edge''.
If the edge has a bend, then the gadget also has a bend.
A visualization of the gadget is given in Figure~\ref{fig:spec_propagation}.

The gadget has two ports representing the endpoints of the edge.
The gadget cannot choose a path for its local curve if both ports receive pressure.
If one port receives pressure, the other gives pressure: as the name suggests, the gadget propagates pressure.
The gadget may also give pressure on both of its ports.

The propagation gadget has a bounding polygon that is a ``thick edge'', having a thickness of 4.
If there is no bend, it is specified as $\langle (0,0) ; (4,0) ; (4, k) ; (0, k) \rangle$ with $k \geq 6$.
If there is a right bend, it is specified as $\langle (0,0) ; (4,0) ; (4, k) ; (4 + k', k) ; (4 + k', k + 4) ; (0, k + 4)\rangle$ with $k, k' \geq 5$.
If there is a left bend, it is specified as $\langle (0,0) ; (4,0) ; (4, k + 4) ; (- k', k + 4) ; (- k', k) ; (0, k)\rangle$ with $k, k' \geq 5$.
We specify coordinates assuming that the left endpoint of the bottom side is at $(0,0)$.
However, the gadget is translation invariant and thus can be placed anywhere for the reduction.

The gates are located at distance 1 from the bottom side, i.e. at $(0,1)$ and at $(4,1)$.
One of the ports is located on the bottom side, from $(0,2)$ to $(0,3)$.
The other port is located on the other side of length 4.
If there is no bend, it is from $(k,2)$ to $(k,3)$.
If there is a right bend, it is from $(4+k', k + 1)$ to $(4+k', k + 2)$.
If there is a left bend, it is from $(-k', k + 2)$ to $(-k', k + 3)$.
On the sides that contain a port, the local network has no edges or vertices (except for those representing the port).
Also, the local curve does not overlap these sides.
This ensures that there is no interaction between gadgets other than via the ports and that the local curves do not intersect.

\begin{figure}[h]
\centering
\includegraphics{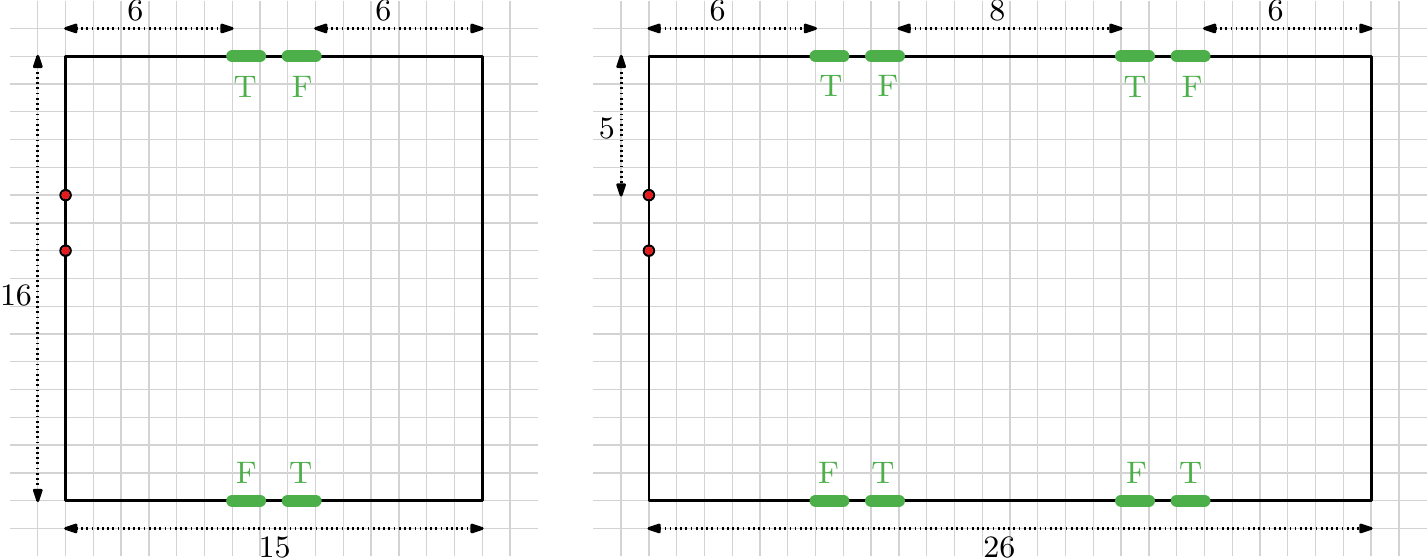}
\caption{Specification of a variable gadget. (left) $k = 1$. (right) $k = 2$.}
\label{fig:spec_variable}
\end{figure}

\paragraph{Variable gadget}
The specification of a variable gadget depends on the number of edges incident to it.
It has zero or more incident edges from above, and zero or more from below.
Let $k$ denote the maximum of the number of incident edges from above and the number of incident edges from below.
We assume that $k > 0$.
A variable with $k = 0$ does not occur in the formula and can be safely omitted.
A visualization of the gadget is given in Figure~\ref{fig:spec_variable}.

This gadget has exactly two choices for its path: one corresponds to the \emph{T-state} (true state), the other to the \emph{F-state} (false state).
It has a number of multiple ports along its top and bottom boundary that give or receive pressure depending on the state.
A \emph{T-port} (true port) gives pressure in the T-state and may receive pressure in the F-state.
Similarly, a \emph{F-port} (false port) gives pressure in the F-state and may receive pressure in the T-state.
For each occurrence (each edge in the embedded formula), the variable gadget has one T-port and one F-port; these are adjacent.

The bounding polygon is a rectangle that depends on the value of $k$.
It has a width of $4 + 11 \cdot k$.
The height is fixed at $16$.

The gates are both located on the left side, at a distance $5$ and $7$ from the top side.
The T- and F-ports are located on the top and bottom boundary of the bounding polygon.
They occur in pairs, one T-port and one F-port, with a distance of 1 between the two.
Between two pairs, there is a distance of 8.
The leftmost and rightmost pair have a distance of 6 to the left and right side respectively.
For the pairs on the top side, the T-port is left of the F-port; for the bottom boundary this is reversed.
There are $k$ pairs on the top side, and $k$ pairs on the bottom side.

\begin{wrapfigure}[12]{r}{.25\textwidth}
\centering
\includegraphics{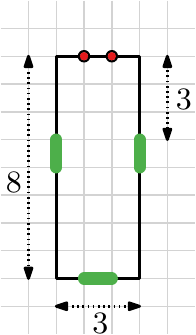}
\caption{Specification of a clause gadget.}
\label{fig:spec_clause}
\end{wrapfigure}
\paragraph{Clause gadget}
The clause gadget has a simple, fixed specification.
A visualization of the gadget is given in Figure~\ref{fig:spec_clause}.

This gadget has three ports.
The clause gadget can choose a path if and only if it does \emph{not} receive pressure on one of its ports.
That is, any path causes on of its ports to give pressure.
The lack of pressure on a port corresponds to the state of the variable being such that the literal is true.
If none of the variables has a state such that the literal is true, i.e. the clause is false, then all ports receive pressure and the clause gadget cannot choose a path.

The bounding polygon is a rectangle, with width 3 and height 8.

The gates are both at the top side and have a distance 1 from the left and right side and a distance 1 between them.
The three ports are placed on the three remaining sides.
The bottom side has a port in the middle.
The left and right side have a port at a distance 3 from the top side.

\paragraph{Construction with gadgets}
With the gadgets defined above, we now construct the complete network and simple closed curve based on the structure given by the (modified) embedded formula $F$.
Figure~\ref{fig:input}(c) illustrates the construction.

First, construct each variable gadget and place these on a horizontal line with a distance of $3$ between consecutive variables.
The order of the variables is given by the embedding of formula $F$.
No variables overlap due to the horizontal spacing.

For the placement of clause gadgets, we introduce \emph{dominance}.
Clause $C$ is said to dominate clause $C'$ if there is a vertical line that intersects both $C$ and $C'$ (or the horizontal part of an incident edge) in $F$ and $C'$ is below $C$.
This induces a (transitive) partial order on the clauses.
Let $\mathop{dom}(C)$ denote the clauses that are dominated by $C$.
The \emph{dominance number} of a clause is $0$ if $\mathop{dom}(C)$ is empty; otherwise, it is $1 + K$ where $K$ is the maximum dominance number of a clause in $\mathop{dom}(C)$.
We can compute the dominance number of all clauses in polynomial time by traversing the given embedding.
We construct the clause gadgets and place them as follows.
Assume the middle literal is the $i$th incident edge from above for the corresponding variable.
If the literal is positive, we place the gadget with its bottom port above the $i$th F-port of the variable.
If the literal is negative, we place the gadget with its bottom port above the $i$th T-port of the variable.
We place the clause with its bottom at height $6 + 12 k$ above the top (or below the bottom) of the variable gadget where $k$ is its dominance number.
Since the height of a clause is 8, the vertical distance between gadgets with a different dominance number is at least 4.
Since $k \geq 0$, the gadgets cannot overlap a variable gadget.
Due to spacing between the ports, no two clause gadgets overlap.

We now construct and place the propagation gadgets.
For each middle edge, we construct the straight propagation gadget of the required length and place it between the clause and the variable.
By the placement rules of the clause, the ports match up correctly: that is, it is connected an F-port if the literal is positive and to a T-port otherwise.
For the outer edges, we construct the propagation gadgets with a bend.
Again, assume the edge is the $i$th incident edge from above for the variable.
We place it such that the bottom port connects to the $i$th F-port if the literal is positive; to the $i$th T-port if the literal is negative.
We give it the proper lengths such that the top port connects to the port of the corresponding clause.
Due to the spacing of the ports, no two vertical parts of a propagation gadget overlap and a propagation gadget does not overlap any variable gadget (except at the bottom port, as desired).
Also, a propagation gadget does not overlap a horizontal part of another propagation gadget nor does it overlap a clause gadget:
such overlaps would contradict the placement via the dominance number.

From the above, we now know how to compose the various gadgets in polynomial time.
We obtain a network, but we do not have a single closed curve: each gadget has its own local (open) curve.
What remains is to argue that we can ``stitch'' the parts together to form a single closed curve.
Since the (order of) the curve is not used to convey information, we may stitch the parts together in any order.
In particular, we proceed as follows to combine the variable gadgets and the clause and propagation gadgets above the variables.
We define a point $s$ that is to the left of the leftmost variable gadget and on the same height as the gates of the propagation gadgets (note that these gates are all on the same height).
We start at $s$ and move to the right.
When we reach a distance of 2 before the left boundary of a variable gadget, we go down to connect to its lower gate, traverse the gadget, go one step left from its upper gate and go back up to the height of $p$.
When we are at a distance of 2 before the left boundary of a propagation gadget that represents the left edge in the embedding (i.e. it has a bend to the right), then we go up and to the right to connect to the right gate of the corresponding clause gadget, traverse it, and go back after exiting from its left gate.
When we reach the gate of a propagation gate, we traverse it and continue from the other gate.
When we reach a distance that is to the right of the right boundary of the rightmost variable gadget, we stop.
We call this point $t$.
Due to the spacing between the gadgets and ports, this traversal does not intersect any gadgets and the described events do not coincide nor do they occur within a propagation gadget.
We add the traversal to the network and use it to define single simple open curve from $s$ to $t$ that visits the traversed gadgets.
For the propagation and clause gadgets below the variables, we use a similar procedure from point $s'$ to $t'$.
The only difference is that the variable gadgets need not be traversed.
By adding the vertical connection between $s$ and $s'$ as well as between $t$ and $t'$ to the curves and network, we obtain a network $G$ and a single simple closed curve $C$.
An example of the traversal is given in Figure~\ref{fig:input}(c).

\paragraph{Proving the theorem}
We now have a network $G$ and a simple closed curve $C$.
Let $n$ denote the number of variables, and $m$ the number of clauses in $F$.
We observe that $6 + n \cdot (4 + 11 \cdot m + 3)$ is an upper bound on the width, and $16 + 6 + 14 \cdot m + 8$ is an upper bound on the height of the grid that is used for the network and curve.
Hence, the space occupied is $O(n m^2)$.
This bounds the size of a propagation gadget and thus the reduction runs in polynomial time.
To prove Theorem~\ref{thm:main}, we must argue that $G$ has a simple cycle $P$ with $\df(C,P) \leq 1$ if and only if $F$ is satisfiable.

Assume that $F$ is satisfiable and consider some satisfying assignment.
We must now argue the existence of a simple cycle $P$.
For each variable gadget, we choose the local path that corresponds to the T-state for a variable with the value true; we choose the local path that corresponds to the F-state otherwise.
This gives pressure on a number of propagation gadgets, and we choose the path for these such that the clause on the other end receives pressure on its port.
For the other propagation gadgets, we choose the path such that it may receive pressure at the clause (and give pressure at the variable).
Since the truth values of the variables originate from a satisfying assignment, we know that at most two ports of any clause receive pressure.
Hence, by the specification, the clause can choose a local simple path as well.
We concatenate the local paths with the paths that are used to stitch together the local curves of various gadgets to obtain a simple cycle $P$.
By construction, $\df(C,P)$ is at most 1.

Now, assume that $G$ contains a simple path $P$ with $\df(C,P) \leq 1$.
This must traverse each variable gadget.
Each path corresponds to either the T- or F-state.
This directly yields the truth values of the variables.
Each clause gadget is also traversed and hence one or more of its ports give pressure.
This means that this also gives pressure to the variable gadget.
This pressure ensures that for this variable is in a state such that that clause is satisfied.
Hence, the truth values give rise to a satisfying assignment.

This proves the theorem.
We have worked only with the specification of the gadgets.
In the next section, we implement the gadgets according to these specifications.

\subsection{Gadget details}\label{ssec:gadgets}

In the previous section, we have proven Theorem~\ref{thm:main} with the specifications of three gadgets.
In this section, we show the how to construct the local network and local curve to meet these specifications.

We illustrate each of the constructions.
We give the local network in thick gray lines and the local curve as a red line, both placed on top of the visual representation of a gadget, given in the previous section.
We also give various path choices, these are given in blue.
A choice of path leads to pressure on the ports.
Ports that give pressure are indicated with an outward arrow.
The thick green lines for ports are also used to indicate edges of intermediate construction blocks that are shared with parts that are not visible in the illustration.

\begin{figure}[h]
\centering
\includegraphics{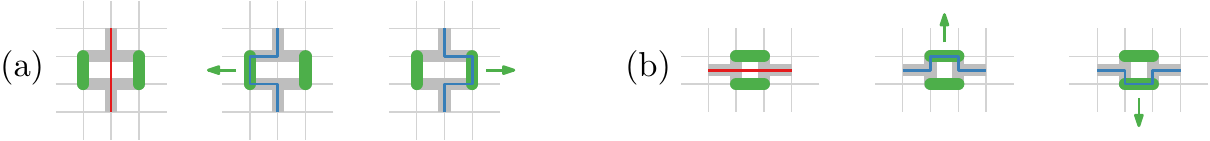}
\caption{Building blocks used to construct a propagation gadget. The local network and local curve are given in thick gray and thin red respectively. Possible paths are given in blue. (a) A horizontal 2-block. (b) A vertical 1-block.}
\label{fig:prop_block}
\end{figure}

\paragraph{Propagation gadget}
To construct a propagation gadget, we use a few simple building blocks.
A \emph{horizontal $w$-block} defines a local network that consists of a rectangle of width $w$ and height 1 and has two \emph{arms} of length 1 protruding from the middle of the horizontal sides.
Analogously, we define a \emph{vertical $h$-block}, with a rectangle of height $h$ and width 1, with two arms protruding from vertical sides.
The curve for the block is a straight line between the far endpoints of the two arms.
We use only 1- and 2-blocks.
These blocks have exactly two possible paths with \f distance at most 1 to the curve.
The choice of path leads to the block giving pressure on one of its sides.
This is illustrated in Figure~\ref{fig:prop_block}.

\begin{figure}[t]
\centering
\includegraphics{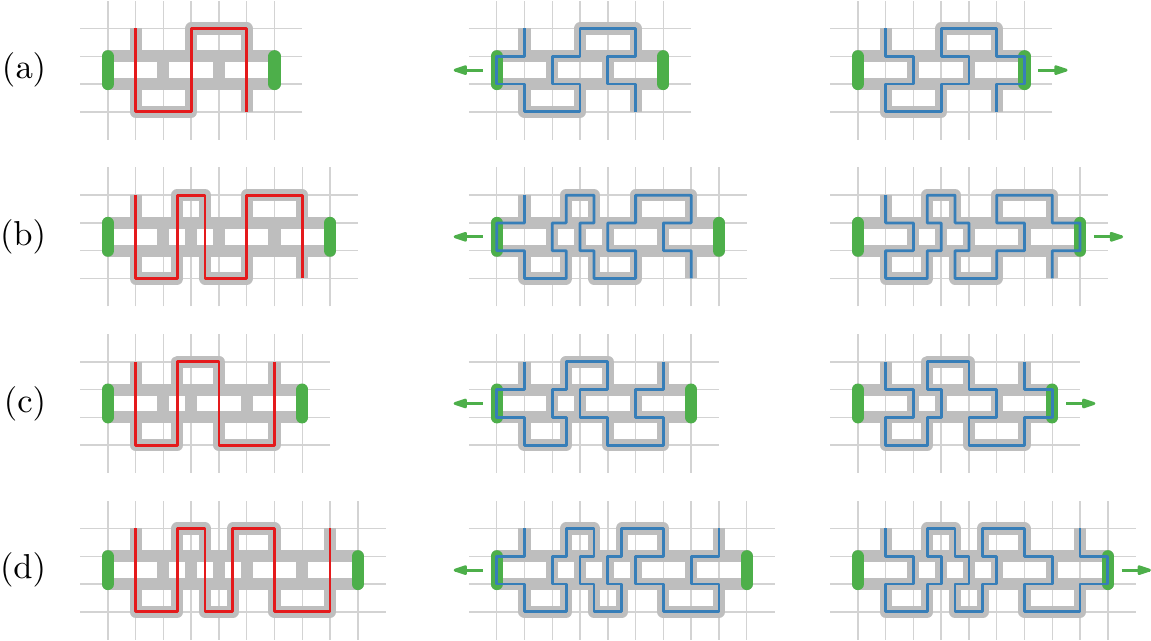}
\caption{Chains of various length and parity requirement. (a) Odd parity with $d = 6$. (b) Odd parity with $d = 8$. (c) Even parity with $d = 7$. (d) Even parity with $d = 9$.}
\label{fig:prop_chain}
\end{figure}

Now we wish to propagate pressure over a horizontal distance $d$ with $d \geq 6$.
To this end, we concatenate a number of horizonal blocks, sharing the vertical sides of the rectangle.
This results in a \emph{horizontal chain}.
Such a chain has a required parity, indicating whether the number of blocks used should be even or odd.
The exact construction depends on the parity of $d$ and whether the parity of $\lfloor \frac{d}{2} \rfloor$ matches the required parity.
Refer to Figure~\ref{fig:prop_chain} for examples.
If $d$ is even and the parities match, we use $\frac{d}{2}$ 2-blocks.
If $d$ is even and the parities do not match, we use a 2-block, followed by two 1-blocks, followed by $\frac{d}{2} - 2$ 2-blocks.
If $d$ is odd and the parities match, we use a 2-block, followed by a 1-block, followed by $\frac{d-1}{2} - 1$ 2-blocks.
If $d$ is odd and the parities do not match, we use a 2-block, followed by three 1-blocks, followed by $\frac{d-1}{2} - 2$ 2-blocks.
The adjacent blocks are joined by the endpoints of the arms, alternatingly above and below the rectangles.
We can choose the alternation freely, i.e. whether the left endpoint of the curve is above or below.
In each of the cases, we see that the chain starts with a 2-block and ends with a 2-block.
Similarly, we construct \emph{vertical chains}.
By construction, it is not possible to choose a path if both ends of a chain receive pressure.
Paths exists that result in a chain giving pressure at both of its endpoints.
These paths have no influence on the reduction and are omitted from the illustrations.

\begin{figure}[t]
\centering
\includegraphics{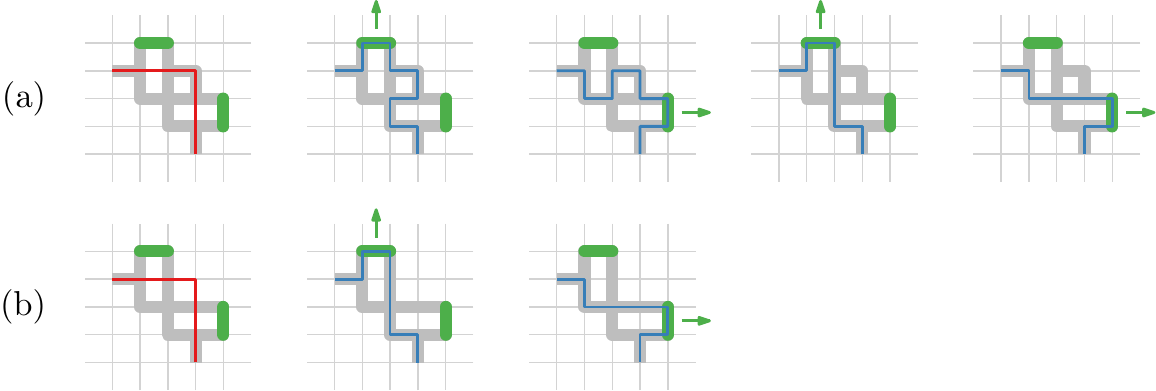}
\caption{A bend is made by combining a horizontal and vertical 2-block. (a) A bend construction with four possible paths. (b) The reduced construction with two possible paths.}
\label{fig:prop_bend}
\end{figure}

To make a bend we combine a horizontal and a vertical 2-block such that they share a vertex.
We require that the curve of the bend starts and ends at the far side of a block with respect to the other block.
This is illustrated in Figure~\ref{fig:prop_bend}(a).
The network edges of the arms on the near sides need not be used by a path; this has no effect on the propagation of pressure.
To obtain exactly two choices, we omit these arms and use the bend construction given in Figure~\ref{fig:prop_bend}(b).
By construction, it is not possible to choose a path if both ends of a bend receive pressure.

\begin{figure}[t]
\centering
\includegraphics{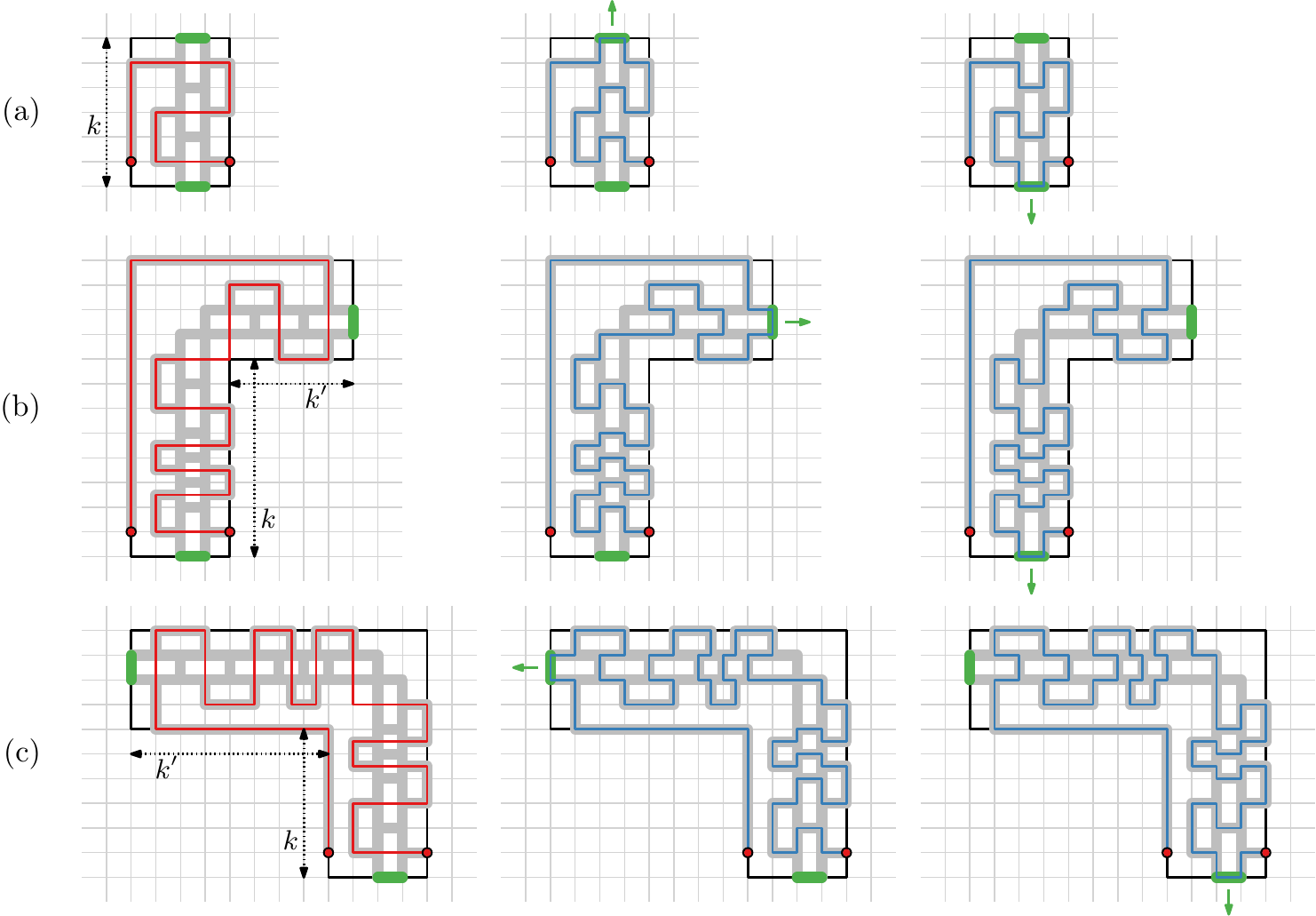}
\caption{Construction of the local network and local curve for a propagation gadget. (a) Without a bend. (b) With a right bend. (c) With a left bend.}
\label{fig:prop_gadget}
\end{figure}

Now we have the required pieces to construct a propagation gadget.
If the gadget has no bend and length $k \geq 6$, we construct a chain of that spans this distance with with odd parity (i.e. with an odd number of blocks).
We choose the alternation such that the bottom end of the curve is on the right; the top end is on the left of the chain.
We place the chain such that the bottom end of the curve corresponds to the right gate.
At the top end, we go one space to the left and then straight down again to the left gate, adding this to the local network and local curve (See Figure~\ref{fig:prop_gadget}(a)).

If the gadget has a right bend, a vertical length of $k \geq 5$, and horizontal length of $k' \geq 5$, we proceed as follows.
We construct a vertical chain of length $k+1$ with even parity and a horizontal chain of length $k' $ with odd parity.
We choose the alternation such that the bottom end of the curve for the vertical chain is on the right; therefore the top end is on the right as well.
We place the vertical chain such that the bottom end of the curve corresponds to the right gate.
We place the horizontal chain, such that its leftmost block and the top block of the vertical chain form a bend to the right.
The alternation is chosen accordingly.
Since the parity is odd, the curve then ends above the chain.
We go one space upwards, then back along bounding polygon of the gadget to the left gate.
This is illustrated in Figure~\ref{fig:prop_gadget}(b).
If the gadget has a left bend, the construction is similar.
However, now we use a vertical chain of length $k+2$ with odd parity and a horizontal chain of length $k'+2$ with even parity.
This is illustrated in Figure~\ref{fig:prop_gadget}(c).

\begin{figure}[b]
\centering
\includegraphics{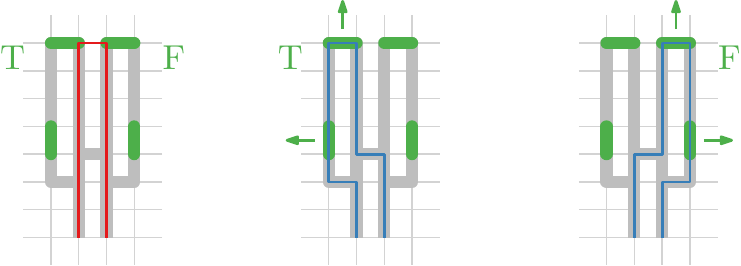}
\caption{An occurrence block has two choices, one for each tine of the ``tuning fork''.}
\label{fig:var_occurrence}
\end{figure}

\begin{figure}[b]
\centering
\includegraphics{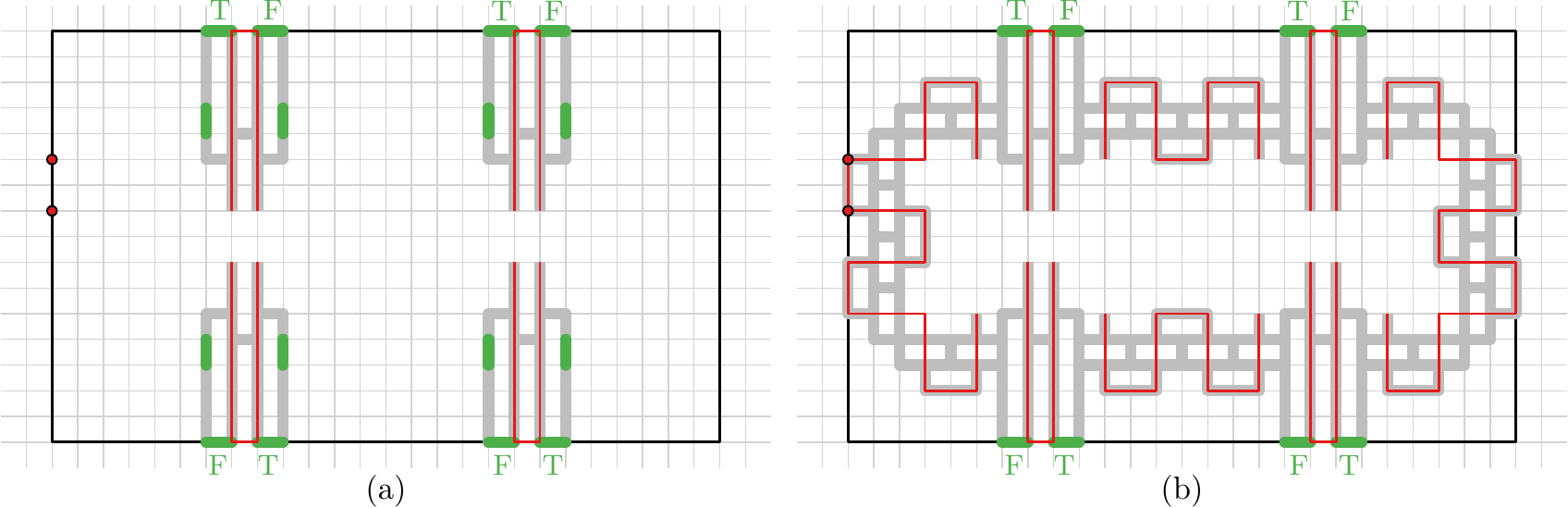}
\caption{(a) Occurrence blocks in a variable gadget. (b) Using chains for a consistent state.}
\label{fig:var_placement}
\end{figure}

\paragraph{Variable gadget}
To construct a variable gadget, we reuse the 2-blocks used to construct a propagation gadget.
In addition, we need another building block that we call the \emph{occurrence block}.
The occurrence block is illustrated in Figure~\ref{fig:var_occurrence}.
Its local network is shaped like a ``tuning fork'': it has two \emph{tines} of height 5, both of width 1.
These are a distance of 1 apart.
It has exactly two choices for its path, one for each tine.
The left tine corresponds to a value ``True'' for the variable; the right tine to a value ``False''.
The top end of the tines correspond to the T-ports and F-ports of the variable gadget.
We place the occurrence blocks accordingly within the gadget (see Figure~\ref{fig:var_placement}(a)).

\begin{figure}[b]
\centering
\includegraphics{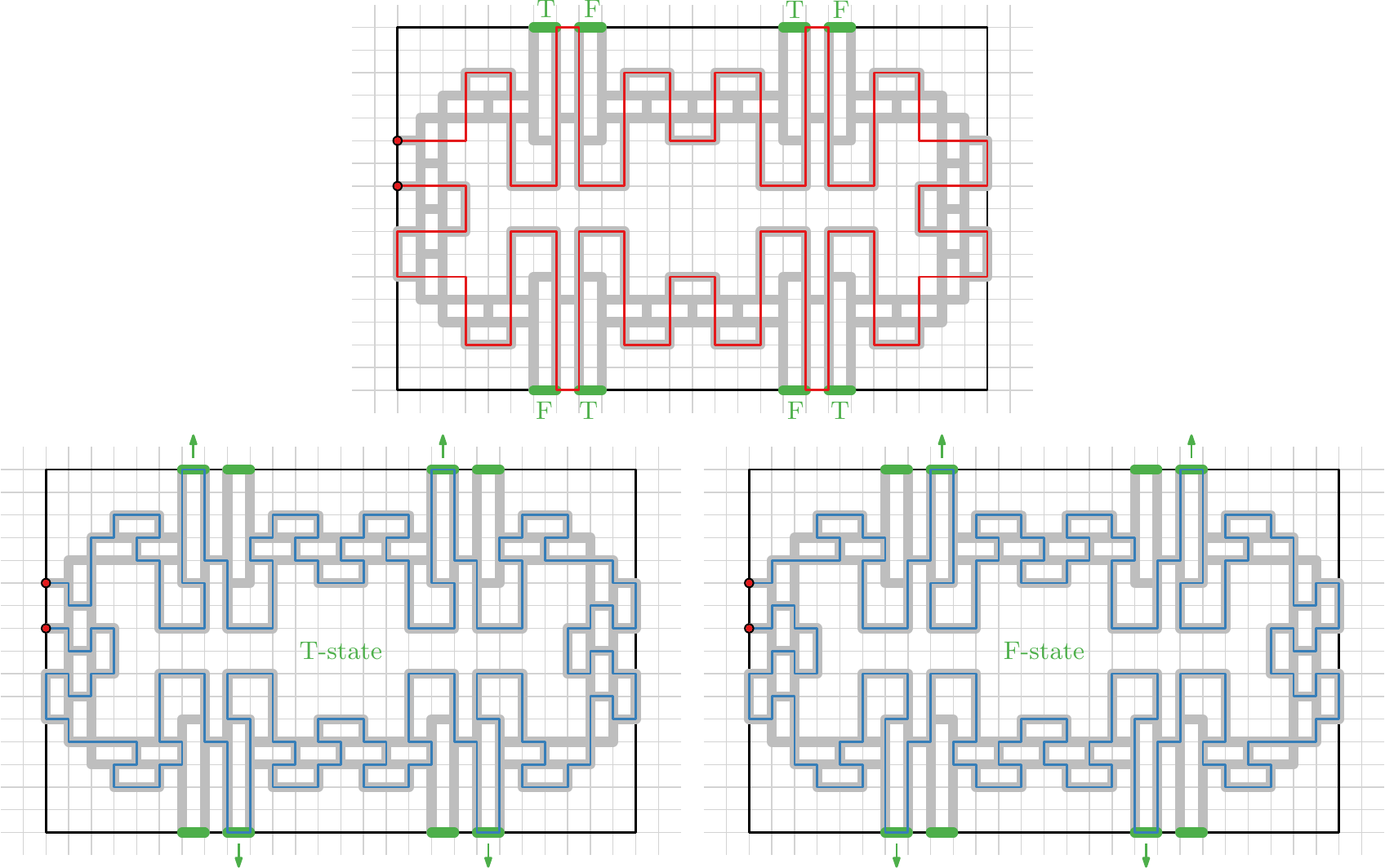}
\caption{The construction of a variable gadget with up to two incident edges from above and from below. The bottom row shows the two states of the gadget.}
\label{fig:var_gadget}
\end{figure}

\newpage
We now need to ensure that all the occurrence blocks choose the same path.
The occurrence blocks must be consistent to have a consistent truth value for the variable.
Using a tine also gives pressure on its outer edge.
Hence, we can use a horizontal chain of length 8 with even parity (that is, four 2-blocks) to connect any two adjacent occurrence blocks.
Moreover, we can use a chain with 2 bends to connect the two leftmost occurrence blocks; analogously, we connect the two rightmost occurrence blocks.
This ensures a consistent state, this is illustrated in Figure~\ref{fig:var_placement}(b).
To obtain a single curve we connect the occurrence blocks and the chains in order.
Moreover, we omit the vertical connection between the two left arms on the left side to adhere to the gates.
The full construction is illustrated in Figure~\ref{fig:var_gadget}.

\begin{figure}[h]
\centering
\includegraphics{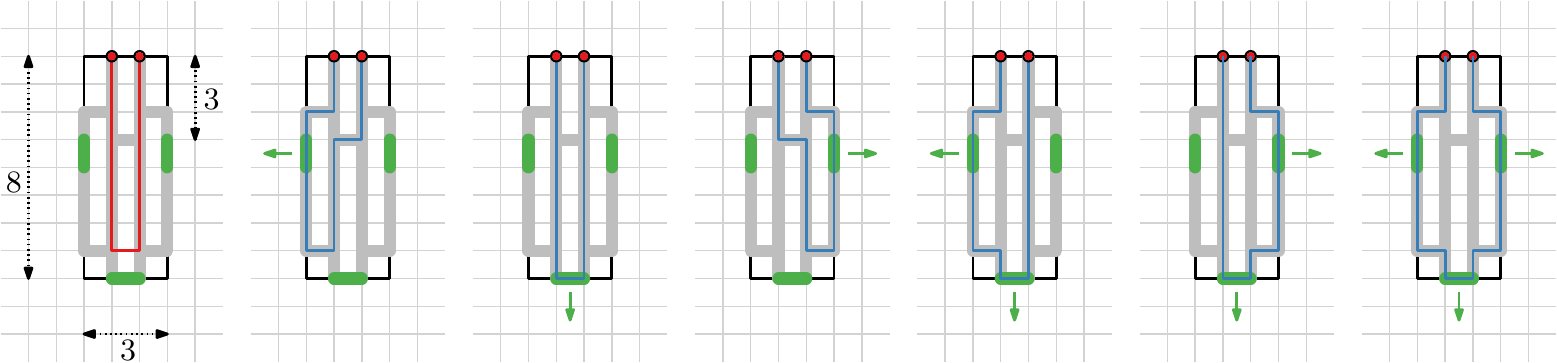}
\caption{A clause gadget with its path choices. Each choice gives pressure on at least one port.}
\label{fig:clause_gadget}
\end{figure}

\paragraph{Clause gadget}
The clause gadget is illustrated in Figure~\ref{fig:clause_gadget}.
It resembles an occurrence gadget: it has been mirrored vertically and gained three edges in the local network at the bottom.
Any choice of a local path results in pressure on at least one port.
Therefore, if all ports receive pressure, no path can be chosen.

\section{Additional results}\label{sec:corollaries}

From Theorem~\ref{thm:main} and its proof, we also obtain a number of related results.

Most importantly, we observe that none of the gadgets can choose a simple path with \f distance greater than 1 and less than $\sqrt{2}$.
Hence, any polynomial-time approximation algorithm with an approximation factor less than $\sqrt{2}$ would be able to decide whether the network $G$ obtained from our construction contains a simple cycle with \f distance at most 1.
\begin{cor}
Let $G$ be a planar straight-line embedding of a graph (a network) and let $C$ be a simple closed curve.
It is NP-hard to approximate the minimal \f distance $\df(C,P)$ of any simple cycle in $G$ within a factor less than $\sqrt{2}$.
\end{cor}
Thus, in particular, no PTAS is possible for this problem.
We observe that there are more path choices for a \f distance of $\sqrt{1.25}$ near the 1-blocks of a propagation gadget.
However, these do not modify the behavior of the gadget: one of its port must give pressure.
For a \f distance of $\sqrt{2}$, a bend (Figure~\ref{fig:prop_bend}) can choose a path that gives pressure on neither of its ports.
This allows propagation gadgets to receive pressure on both gates and thus they would no longer adhere to the specification.
These bends are also used to obtain a consistent state within a variable gadget.
As a result of the bend no longer functioning, ports at the top and bottom may have different states.

Moreover, we may also observe that all interaction between and within gadgets is based on edges.
Hence, from the same construction, we know that it is NP-hard to determine the existence of an ``edge-simple'' cycle that uses each edge at most once.
There is in fact one exception: the construction of a bend (Figure~\ref{fig:prop_bend}).
However, the possible choices for a path remain the same.

The proof straightforwardly extends to open curves and paths in the network.
Omitting the connection between $s$ and $s'$ (see Figure~\ref{fig:input}) yields a simple open curve, rather than a closed curve.
Also, generalizations of the problem in which the network is not a planar embedding or the curve is not simple are NP-complete.

We also observe that the monotonicity of the \f distance is not crucial in our proof.
Hence, the problem under the weak \f distance is also NP-complete.
Using the weak \f distance does allow for some more paths in the local networks.
The occurrence block now has a path choice that uses both ``tines''.
However, this choice cannot be used due to the circular pressure within the variable gadget.
In addition, the clause gadget now has a path choice that results in pressure on both the left and right port (but not the bottom port); this choice adheres to the specification.
Hence, this change does not influence the proof.

\section{Conclusions}\label{sec:conclusion}

We have shown that it is NP-complete to determine the existence of a simple cycle $P$ in a network such that the \f distance of $P$ to some given simple closed curve is at most 1.
In addition, we considered the implications of our proof an a number of variants of this problem.

\paragraph{Future work}
An interesting question is to see what further restrictions we can pose on the input.
For example, the constructed network is relatively sparse and constructed specifically for the proof.
What if we restrict our choice in network, e.g. is the problem NP-complete if we restrict the network to a regular square grid?

\paragraph{Acknowledgements}
I would like to thank Kevin Buchin and Bettina Speckmann for reviewing draft versions and helping me with the presentation of this paper.

\bibliographystyle{abbrv}

\end{document}